\documentstyle[PASJadd,epsfig]{PASJ95}
\newcommand{\vol}{53}
\newcommand{\no}{2}
\newcommand{\lett}{}
\newcommand{\spage}{in press; number of pages}
\newcommand{\rdate}{2000 December 19}
\newcommand{\adate}{2001 January 29}
\newcommand{\beq}{\begin{equation}}
\newcommand{\eeq}{\end{equation}}

\newcommand{\dg}{^{\rm{o}}}
\newcommand{\thetao}{{\theta_{\rm{o}}}}

\newcommand{\varphie}{{\varphi_{\rm{e}}}}
\newcommand{\nuo}{{\nu_{\rm{o}}}}

\newcommand{\nue}{{\nu_{\rm{e}}}}
\newcommand{\rme}{{r_{\rm{e}}}}

\newcommand{\astar}{a}
\newcommand{\bmath}[1]{\mbox{\boldmath{${#1}$}}}

\newcommand{\rms}{{r_{\rm ms}}}

\newcommand{\rg}{{r_{\rm g}}}
\newcommand{\lag}{{\delta t}}
\newcommand{\lagmax}{{\delta t_{\rm m}}}

\def\spose#1{\hbox to 0pt{#1\hss}}
\def\ltasim{\mathrel{\spose{\lower 3pt\hbox{$\mathchar"218$}}
 \raise 2.0pt\hbox{$\mathchar"13C$}}}
\def\gtasim{\mathrel{\spose{\lower 3pt\hbox{$\mathchar"218$}}
 \raise 2.0pt\hbox{$\mathchar"13E$}}}

\newcounter{fA}
\newcounter{fB}
\newcounter{fC}
\newcounter{fD}
\newcounter{fE}
\newcounter{fF}
\newcounter{fG}

\begin{document}
\setcounter{page}{001}

\title{Variable Line Profiles Due to Non-Axisymmetric Patterns
        in an Accretion Disc around a Rotating Black Hole$^1$}

\author{Vladim\'{\i}r {\sc{}Karas},$\!^{\ast}$
        Andrea {\sc{}Martocchia},$\!^{\dag,\ddag}$ and
        Ladislav {\sc{}\v{S}ubr}$^{\ast}$\\[16pt]
$^{\ast}$~{\it{}Astronomical Institute, Charles University Prague,}
 {\it{}V~Hole\v{s}ovi\v{c}k\'ach~2, CZ-180\,00~Praha, Czech Republic}\\
$^{\dag}$~{\it{}Scuola Internazionale Superiore di Studi Avanzati,}
 {\it{}Via Beirut~2-4, I-34\,014~Trieste, Italy}\\
$^{\ddag}$~{\it{}Terza Universit\`a degli studi di Roma,}
 {\it{}Dipartimento di Fisica, Via della Vasca Navale~84, 
 I-00\,146 Roma, Italy}\\[4pt]
 {\it{}vladimir.karas@mff.cuni.cz,
       martok@haendel.fis.uniroma3.it,
       subr@aglaja.ms.mff.cuni.cz}
}

\abst{We have explored spectral line profiles due to spiral patterns in
accretion discs around black holes. A parametrization was employed
for the shape and emissivity of spiral waves, which can be produced by
non-axisymmetric perturbations affecting the disc density and ionization
structure. The effects of the light-travel time, energy shift, and 
gravitational focusing near to a rotating black hole were taken into account. 
A high-resolution ray-tracing code was used to follow the time variations of
the synthetic line profile. A variety of expected spectral features were
examined and the scheme applied to a broad iron line observed
in MCG$-$6--30--15.}

\kword{accretion, accretion disks -- black hole physics -- galaxies: nuclei 
 -- galaxies: individual (MCG$-$6--30--15) -- X-rays: galaxies}

\maketitle

\stepcounter{footnote}
\footnotetext{Low-resolution figures have been used in order to conform to the 
standards of {\sf{}astro-ph} e-print library.}

%%%%%%%%%%%%%%%%%%%%%%%%%%%%%%%%%%%%%%%%%%%%%%%%%%%%%%%%%%%%%%%%%%%%%
\section{Introduction}
Time-dependent phenomena are of extreme richness, variety,
and observational relevance for accreting black-hole systems (e.g.,
Kato et al.\ 1998; Krolik 1999). Clearly, astrophysically
realistic processes must result from the interplay of different factors
which affect light-curves and produce spectral features. Here we
concentrate on the X-ray spectral band.

Rapid temporal changes of the flux and of individual spectral features
are observed in many active galactic nuclei (AGN) and in Galactic
black-hole candidates (GBHCs). Variability patterns fit well with the
widely adopted scenario of these objects as massive black holes
surrounded by accretion flows, in the form of a disc or a torus (e.g.,
Ulrich et al.\ 1997). The X-ray variability of AGN and of
GBHCs is often attributed to instabilities in their accretion flows
causing changes in the inner and outer boundaries of the radiating disc
region or relatively persistent non-axisymmetric inhomogeneities in the
disc itself. This paper deals with variability properties which can be
due to spiral structures forming in the inner part of the disc at
distances $<100\rg$ from the centre. We consider rotating (Kerr) black
holes together with spiral waves of density perturbation in the disc
matter. The waves are supposed to modify the emissivity pattern of the
gaseous flow.

A power-law continuum is observed in X-ray spectra, which is thought to
provide primary irradiation of the flow. Numerical computations of local
reflection spectra from X-ray irradiated material suggest that the
emissivity depends strongly on the nature and location of the primary
source (e.g., Martocchia, Matt 1996; Martocchia 2000, and references
therein), and also on the ionization structure and density profile of
the flow (Done, Nayakshin 2001). Only in the case of density larger than a
critical value and small ionization parameter, the matter behaves
like a pure optically thick medium from which a typical cold iron
line is emitted (Nayakshin et al.\ 2000; Dumont et al.\
2000). The extent to which ionization in the disk influences 
emission and absorption features is subject of much debate.

As far as the power-law continuum is concerned, a plausible scheme
suggests that this spectral component is produced in a hot Comptonizing
medium (corona) above the disc. Another possibility is that magnetic
reconnection causes temporary, strong flashes of radiation (flares) near
the inflowing matter (Galeev et al.\ 1979; Coppi 1992; Haardt et al.\
1994; Nayakshin 1998). Several mechanisms have
been proposed for flares, but the available spectroscopy does not
allow us to decide among them; see Malzac and Jourdain (2000) for recent
discussion and further references. The flares are supposed to irradiate
the gas in their neighbourhood, contributing to short-term variability
(Mineshige et al.\ 1994; Yonehara et al.\ 1997).
Furthermore, the coronal activity may be localized in asymmetric and
rapidly evolving areas. In both cases (flares vs.\ inhomogeneous
coronal activity) variations in the primary flux are expected, which
also determine the variation in the additional spectral features due to
Compton reflection (a continuum ``hump'') and fluorescence (lines).

In this work we explored the expected spectral line profiles in the
region where relativistic light-travel time, energy shift and lensing
effects influence substantially the observed radiation fluxes, temporal
variability, and the lines centroid energy. We assumed that a
perturbation (of density and ionization structure) develops on 
length-scales of ten or a few tens of gravitational radii, giving rise
to emissivity contrast against the background of the axisymmetric flow.
The spirals become active after being illuminated from the primary source.
Here, one has to distinguish between two cases: (i)~the primary
illumination independent from the origin of the spiral perturbation of
the disc; (ii)~the two phenomena (the disc perturbation and the flare)
being connected to each other. The second case is particularly relevant
for binary systems with Roche-lobe overflow; in such a situation,
the outer end of the spiral pattern coincides with the place where the
overflowing material hits the disc. This is the place where a flare or a
spot develops. Therefore, we consider the emissivity variations with
simultaneous contributions from the spiral pattern and from the flare.

We computed the predicted, variable line profiles with the aim of
assessing the importance of the mentioned effects, especially the degree
of the line flux and energy fluctuations that could be assigned to the
existence of such non-axial structures. We also mention our
computational approach, which can be adapted to different intrinsic
emissivities and the source geometry, and, for the sake of illustration, 
we determine free parameters of the phenomenological model for 
the variable iron K$\alpha$ emission of the Seyfert galaxy 
MCG$-$6--30--15.

Rapid variations have been reported in the X-ray continuum and in the
K$\alpha$ iron line of MCG$-$6--30--15 (6--7~keV), namely, Iwasawa et al.\
(1996, 1999) described the spectral changes which occurred during 
ASCA observations in 1994 and 1997. It is important to examine how
the changes of selected spectral features --- and the fluorescent iron
feature in particular --- reflect the changes in primary illumination,
because this helps to constrain the geometry of the emitting regions, for
instance by reverberation techniques (Reynolds et al.\ 1999; for a
recent review, see Fabian et al.\ 2000).

We start the next section by giving some further hints on the physical
models which justify our interest in considering the spiral-type
patterns.

\section{The Model}%---------------------------------------------------
In AGN, variability time-scales go down to less than one hour,
which is almost comparable with the light-crossing time ($t_{\rm{c}}$) at
distances of the order of the gravitational radius ($\rg$) of a
supermassive black hole: $t_{\rm{c}}\approx10^2M_7$\,s, where $M_7$ is
the central mass in units of $10^7M_\odot$. Other time-scales relevant to
black-hole accretion discs (orbital, thermal, sound-crossing, and
viscous) are typically longer than $t_{\rm{}c}$. Characteristic
time-scales of the observed variability also depend on the inclination of
the system with respect to an observer, and can be scaled with the
mass of the central object. Moreover, time intervals become longer very
near to the hole if gravitational delays predicted by general relativity
are taken into consideration.

\subsection{Astrophysical Motivations}%---------------------------------
The irradiation of the disc from a varying source of primary
X-rays does not lead to instantaneous response, but generates
reflection or fluorescence. A delay occurs depending on the geometrical and 
physical state of the gaseous material. For example, Blandford and McKee (1982)
and Taylor (1996) have discussed the intrinsic delays of the response in the
case of clouds geometry, but here such effects are not considered for
simplicity. Instead, it is the location of the primary source ---
often identified in one or more flaring regions --- together with
reflection patterns in the disc which determine the observed signal. A
single variability event is therefore made of a primary flare and a
complex response due to Compton reflection (or fluorescence, in the case
of the iron line) by the disc matter at various distances from the hole.

Several people have considered the reprocessing of primary X-rays recently 
(for references, see Dumont et al.\ 2000; Karas et al.\ 2000; Nayakshin et
al.\ 2000). Reprocessed radiation reaches the observer from
different regions of the system. Individual rays experience
unequal time lags of $\lag$ for purely geometrical reasons, and other
relativistic effects occur in the very vicinity of the hole and are
relevant for the source variability, namely, lensing and Doppler
boosting (Rauch, Blandford 1994). In the case of long-lived patterns 
($t\gg{}t_{\rm{c}}$) in the disc, there is, indeed, a substantial 
contribution to the variability caused by the orbital motion
(Abramowicz et al.\ 1991; Mangalam, Wiita 1993). If the disc is
strictly equatorial, two types of such patterns have been examined in
more detail: spots (which could be identified with vortices in gaseous
discs; cf.\ Abramowicz et al.\ 1992; Adams, Watkins 1995; Karas 1997),
and spiral waves. In principle it is possible to distinguish
different origins of the varying emission, namely, short-lived flares
which decay on the dynamical time-scale versus long-duration features
(Lawrence, Papadakis 1993; Kawaguchi et al.\ 2000). To that purpose,
the geometry of the system needs to be constrained sufficiently. In
addition to intrinsic variations, obscuration is involved along the path
of light rays towards the observer. The light propagation is affected by
the presence of refractive interstellar medium on the line of sight
(Rickett et al.\ 1984) which contributes, together with
internal and random obscuration, to rapid and high-amplitude flickering
seen by the distant observer (Abrassart, Czerny 2000). These
non-intrinsic effects are beyond the scope of this paper. 

The spiral
waves represent large-scale structures (size comparable with the radius),
which can be induced by non-axisymmetric perturbations (Sanders,
Huntley 1976; Wada 1994; Lee, Goodman 2000) or due to magnetic
instabilities (Tagger et al.\ 1990). Also, a pattern resembling a
single-armed spiral is produced from an extended spot after its decay
due to shear motion in the differentially rotating disc. Even though 
such a pattern is only a transient feature, it may last sufficiently long
to produce observable effects. Thus, the spirals are expected to arise
naturally in accretion discs and to persist there for times longer than
the orbital period of the flow particles.
Sanders, Huntley (1976) demonstrated that the formation of a two-arm
spiral wave is a natural response of differentially rotating gaseous
discs to an oval gravitational perturbation, much like in the case of a
self-gravitating stellar system into which the gas is assumed to be
embedded. Recently, Kuo and Yuan (1999) examined the excitation
of spiral waves in Lindblad resonances by comparing the asymptotic
theory of bar-driven waves with the corresponding results of numerical
computations. In a different way, Tagger and Pellat (1999) showed that
spirals can be formed due to a certain kind of magnetic instability;
such waves remain localized close to the inner radius of the disc. 
In general, spiral patterns can develop an odd or even number of arms
depending on the type of the perturbing forces that act on the disc.

Spiral patterns have been treated in the theory of stellar systems, and
widely discussed in the context of non-axisymmetric perturbations which
can modify density profiles and emissivities of gas discs (Goldreich,
Tremaine 1979; Sawada et al.\ 1986; R\'o\.zyczka, Spruit
1993). The status of {\sc{}SPH} computations was summarized recently by
Lanzafame, Maravigna and Belvedere (2000) who examined the role of
stellar-mass ratio on the formation of spiral structures in binaries.
The spirals meet numerous applications also with respect to cataclysmic
variables (Steeghs, Stehle 1999). Furthermore, the modulation of X-rays by
spiral waves was also proposed with regard to polars (Murray et al.\
1999).

Most of the attention with regard to spiral-waves spectral features has
been motivated by studies of cataclysmic variables and other
stellar-mass binary systems. It was noted that the variation in the
density profile and ionization of accretion flows, predicted by numerical
and semi-analytical methods, is followed by temperature modulation, and
therefore by a change in the gas (thermal) emissivity within the
spiral patterns. A similar effect is also expected for the X-ray irradiated
accretion flows in AGN due to an increase in the matter density along the
patterns (if optical thickness $\tau\ltasim1$) or variations in the
ionization structure (the emissivity is not sensitive to density changes
if $\tau\gg1$). A phenomenological description appears to be adequate for
the current scenario of X-rays reprocessing in AGN cores, and is
convenient for the numerical computation of line profiles.

As far as AGNs are concerned, spiral waves have been explored as possible
sources of spectral line variations in the context of broad
emission-line regions (Chen, Halpern 1990; Chakrabarti, Wiita 1993,
1994; Eracleous 1998) and in maser sources (Maoz, 
McKee 1998). In these studies, because the radiation originates at distances
$\gtasim10^3\rg$ from the centre, general-relativity effects are
unimportant and variability time-scales are of the order of several
months to years. On the other hand, Sanbuichi, Fukue and Kojima (1994) 
considered spirals extending close to a Schwarzschild black hole where the
effects of general relativity play a role, and showed examples of 
relativistically distorted spectra.

\begin{fv}{\thefA}{0pc \epsfxsize=\hsize\epsfbox{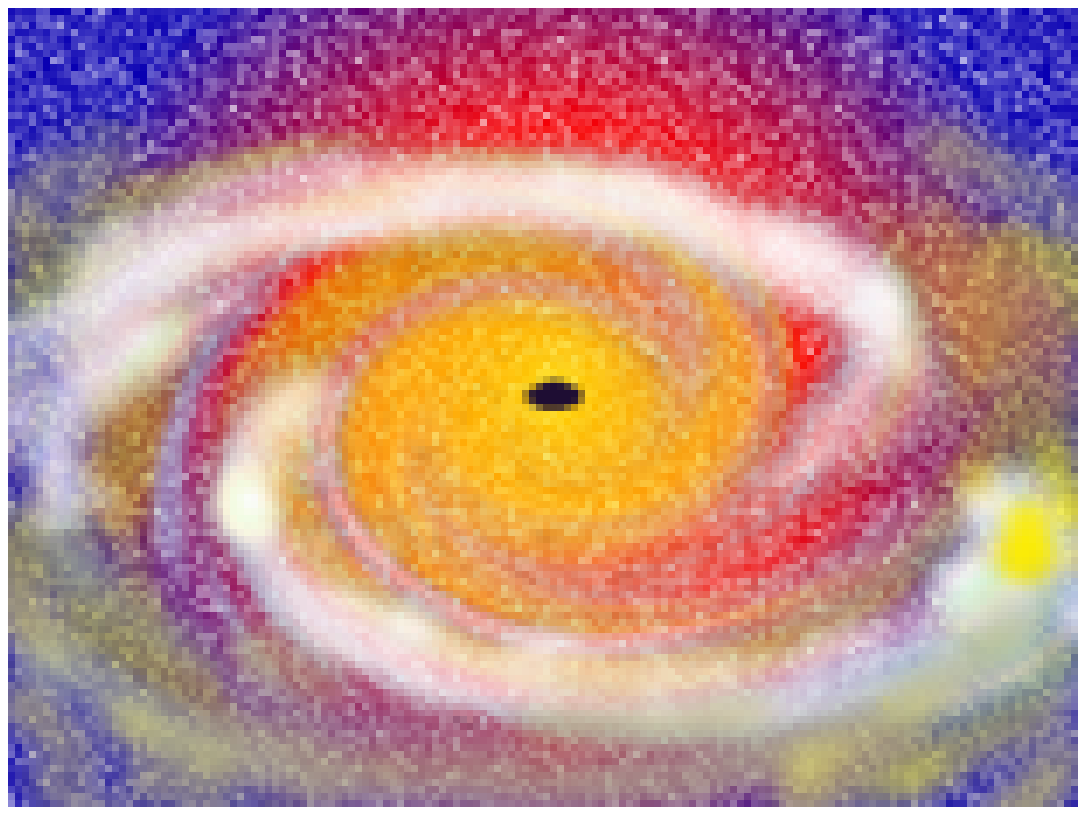}}
{Geometry of the model: an artistic expression of spiral patterns in an
 accretion disc. In the present work, the intrinsic emissivity is
 described by a small number of free parameters which also determine the
 variability of the spectral features. We consider the emissivity to be
 modulated by the spiral pattern with an associated decaying flare.}
\end{fv}%.............................................................

\subsection{A Toy Model of Spiral-Wave Emissivity}%-------------------
We parametrize the Kerr spacetime in terms of the black hole
specific angular momentum $(0\leq\astar\leq1)$. In the disc plane, we
consider both the spiral and a decaying spot localized at the spiral's
outer end (radial location and azimuth of the spot are thus determined 
by the spiral's shape and its orientation with respect to the observer). 
Such a spot can represent a temporary off-axis flare (its maximum amplitude
$A_{\rm{}m}$ is a free parameter). See figure~\thefA\ for a general
scheme of the model.

The complexities of primary X-ray reprocessing are overpassed by
directly parametrizing the emissivity. Our approach to the emissivity
parametrization is close to the computations of Sanbuichi et al.\ (1994)
and Bao and Wiita (1999), who considered a non-rotating black hole.

\begin{fv}{\thefB}{0pc \epsfxsize=\hsize\epsfbox{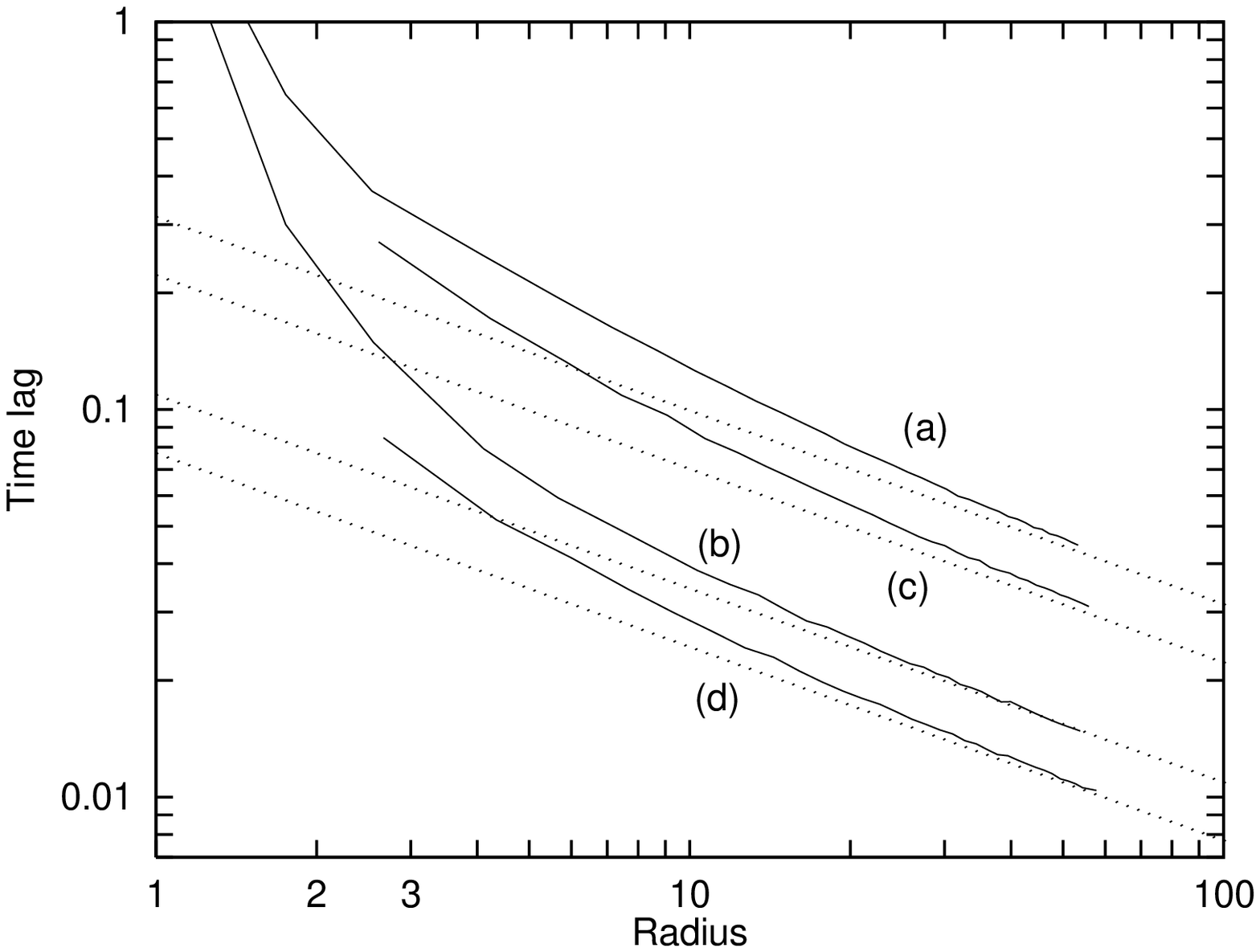}}
{Maximum time lag $\protect\lagmax$ shown between the rays from a
ring of given radius in the Kerr metric (solid curves). Four cases are
plotted for radii greater than $\rms$ and different combinations with
the black-hole rotation parameter, $\protect\astar$. The observer
inclination angle $\protect\thetao$ is: (a)~$\astar=1$,
$\protect\thetao=80\protect\dg$; (b)~$\astar=1$,
$\protect\thetao=20\protect\dg$; (c)~$\astar=0$,
$\protect\thetao=80\protect\dg$; (d)~$\astar=0$,
$\protect\thetao=20\protect\dg$. Radius is given in units of $\rg$,
while the time lag is expressed with respect to Keplerian orbital period,
$t_{\rm{}k}$. The Euclidean estimate of $\protect\lagmax$ is also shown
(dotted lines).}
\end{fv}%..............................................................

The spiral pattern is characterized by the emissivity distribution in
the equatorial plane,
\beq
j_{\rm{e}}^{\rm{}(s)}(\nu;\rme,\varphie)=
 j_0(\nu-\nu_0)\,\rme^{-\gamma}\,\sin^\beta\varphie(\rme),
\label{j0}
\eeq
where the frequency-dependent term, $j_0$, 
is taken as a narrow Gaussian profile,
while the power-law index, $\gamma$, describes the overall radial decrease
in the emissivity. The spiral rotates with respect to the observer, and
its shape is logarithmic and determined by the function
$\varphie(r)=\varphi+\alpha\log(r/r_0)$, where $\arctan\alpha$ is the
pitch angle, and $r_0$ is the outer radius where the wave is being
excited.\footnote{The locally emitted intensity (denoted by subscript
``e'') is related to the observed intensity at spatial infinity (subscript
``o'') by the formula
$j_{\rm{e}}(\nue;t-\lag)/\nu_{\rm{e}}^3=j_{\rm{o}}(\nuo;t)/\nu_{\rm{o}}^3,$
where the ratio $g\equiv\nuo/\nue$ defines the redshift factor and
$\lag\equiv\lag(\rme,\varphie;\thetao,\astar)$ is time delay due to the
light-travel time along the ray. See Fanton et al.\ (1997) and
Martocchia, Karas and Matt (2000) for further details on ray-tracing
computations.} This parametrization approximates the spiral pattern
evolving on the background of a Keplerian disc. The azimuthal form of
the spiral depends also on the type of perturbation, so what we use here
is one of the possible modes, which appears to be simple enough for
illustration purposes.

As already discussed, the model takes inspiration from spirals in close
binaries where a spot arises following the interaction between the
stream (after the Roche-lobe overflow) and the disc (e.g., Wolf et al.\ 1998).
The radiation flux in the spot decays with an e-folding time of $q\gtasim1$
orbital periods, $t_{\rm{}k}=2{\pi}(r_{\rm{e}}^{3/2}+\astar)$. Again,
the frequency-dependent part, $j_0$, of the spot intrinsic profile is taken 
in the Gaussian form and is assumed to be sufficiently narrow (its smearing
is due mainly to radial component of the source orbital motion), while
the amplitude, $A(r,\varphi)$, decreases exponentially with the distance
from the flare center located at $r_0$,
\beq
j_{\rm{e}}^{\rm{}(f)}(\nu;\rme,\varphie)=
 j_0(\nu-\nu_0)\,A(\rme,\varphie)\,t^{-q}.
\label{j1}
\eeq
Here, $A(r,\varphi)=A_{\rm{}m}\exp(-\Delta{r})$ with, typically, the
spot size $\Delta{r}=|\bmath{r}-\bmath{r}_0|\ltasim1$ and
$r_0\approx20$. Hence, the resulting emissivity is a sum of
contributions (\ref{j0}) and (\ref{j1}):
$j_{\rm{e}}=j_{\rm{e}}^{\rm{}(s)}+j_{\rm{e}}^{\rm{}(f)}$.

It is the exact form of the emissivity distribution along the disc
surface which determines the observed flux in the line wings and, in
particular, which of the two wings --- the bluish or the reddish one,
appears more pronounced in the predicted spectrum. Thereby the lines are
more complex than one would expect on the basis of intuition with simple
ring-type sources (double-horn profiles). Here, we assume that a
spiral with two arms extends from $r=r_0$ down to $\rms$, being
terminated at the innermost stable orbit. Single-armed spirals show
comparable spectral characteristics, but their variations occur with a
double period and a somewhat higher amplitude, similar to the case of an
elongated spot.

\begin{figure*}%............................................................
\vspace*{-3mm}%offset
\epsfxsize=0.37\hsize\makebox[0.42\hsize]{\hspace*{10ex}\epsfbox{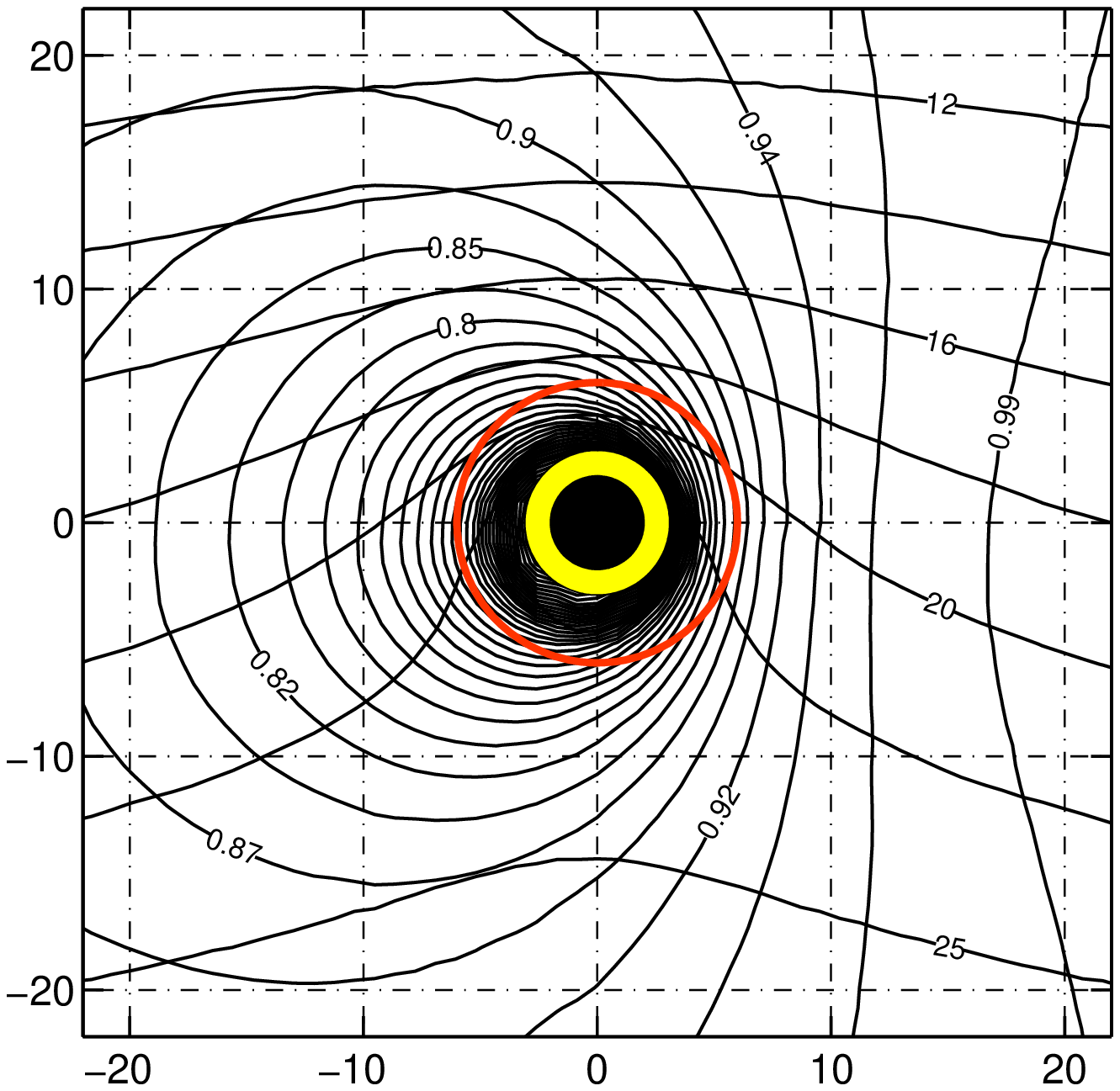}}
\vspace*{3mm}%offset
\hfill
\epsfxsize=0.37\hsize\makebox[0.42\hsize]{\hspace*{-10ex}\epsfbox{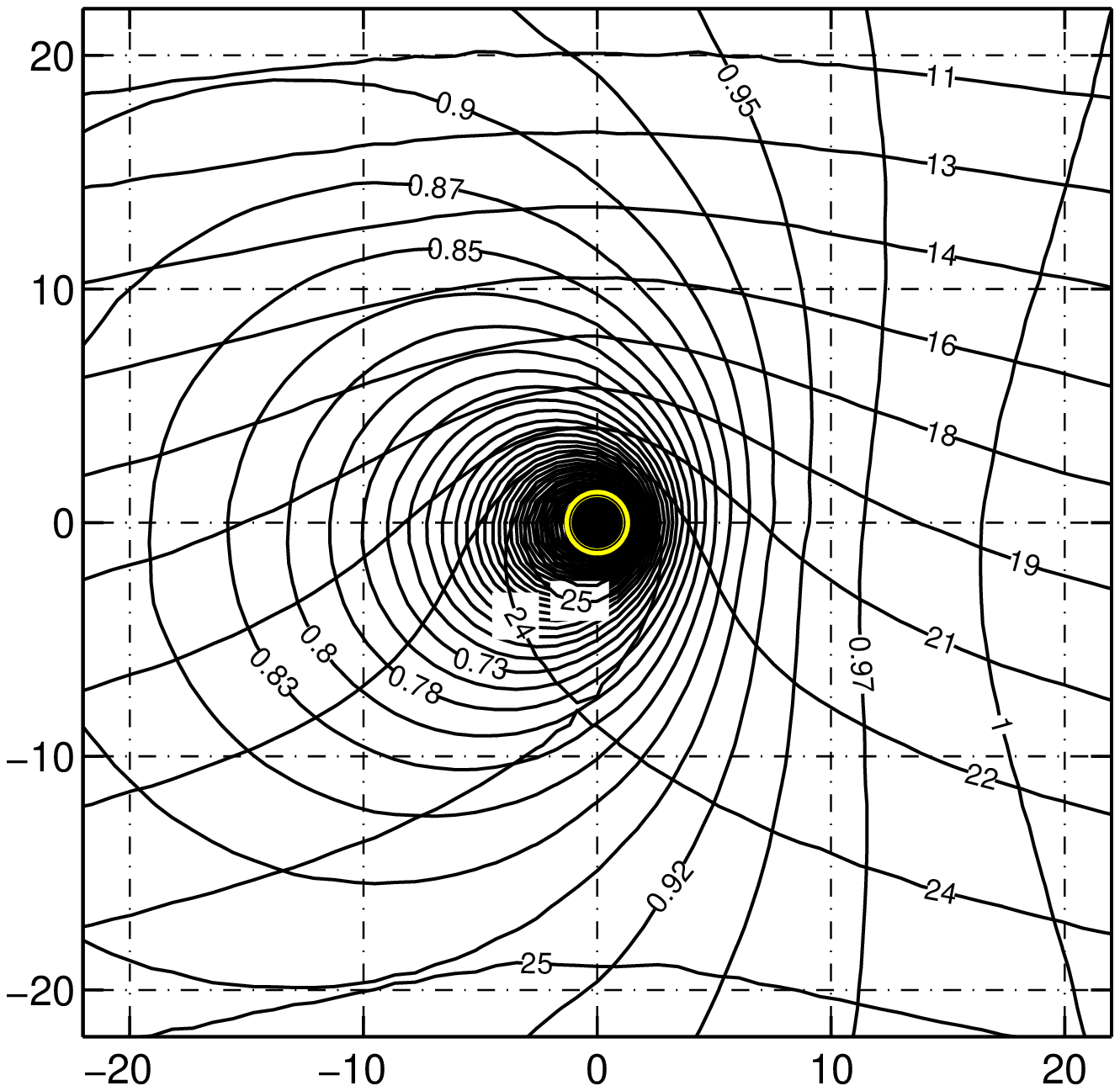}}
\vspace*{3mm}%offset
\epsfxsize=0.37\hsize\makebox[0.42\hsize]{\hspace*{10ex}\epsfbox{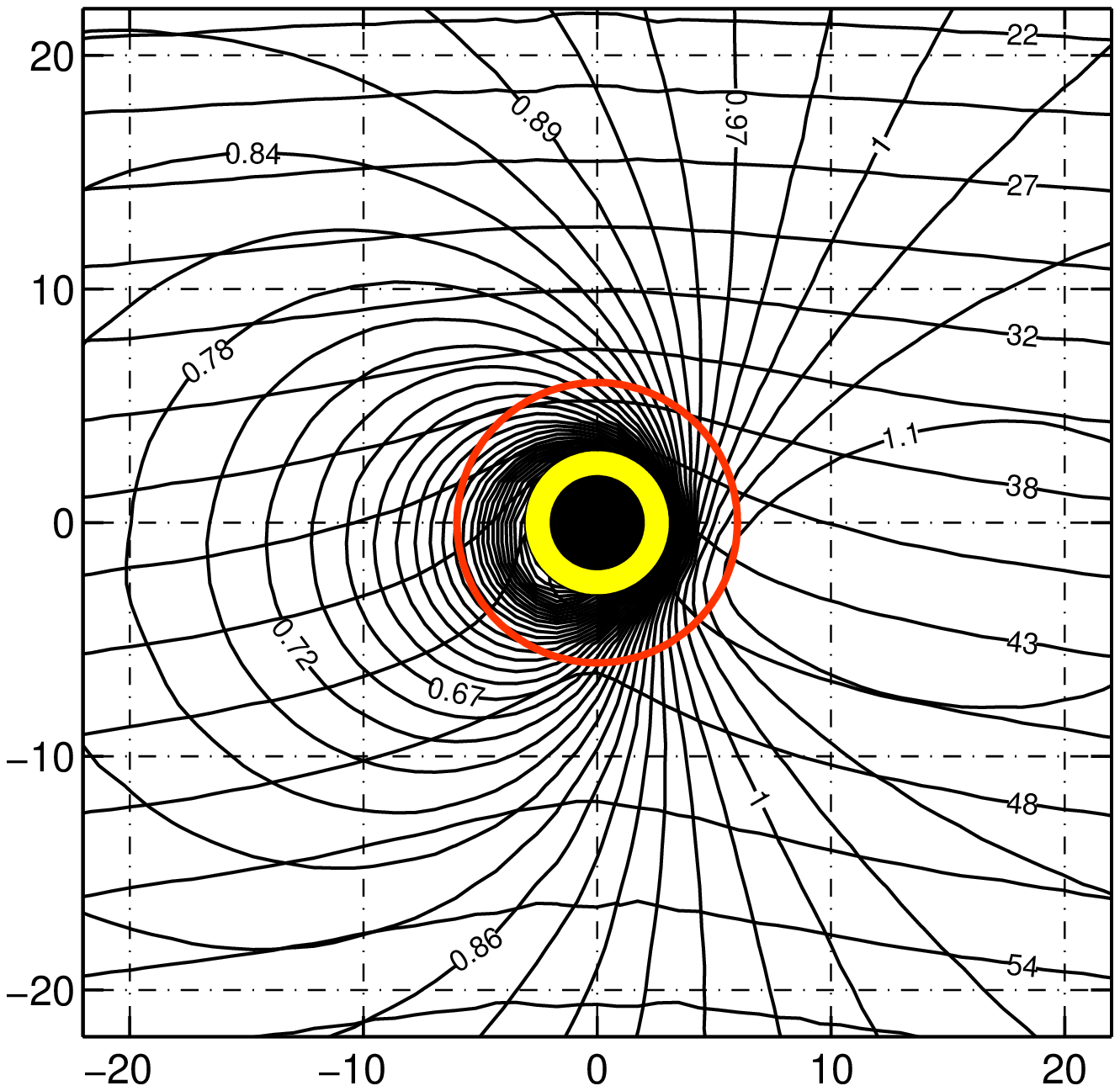}}
\hfill
\epsfxsize=0.37\hsize\makebox[0.42\hsize]{\hspace*{-10ex}\epsfbox{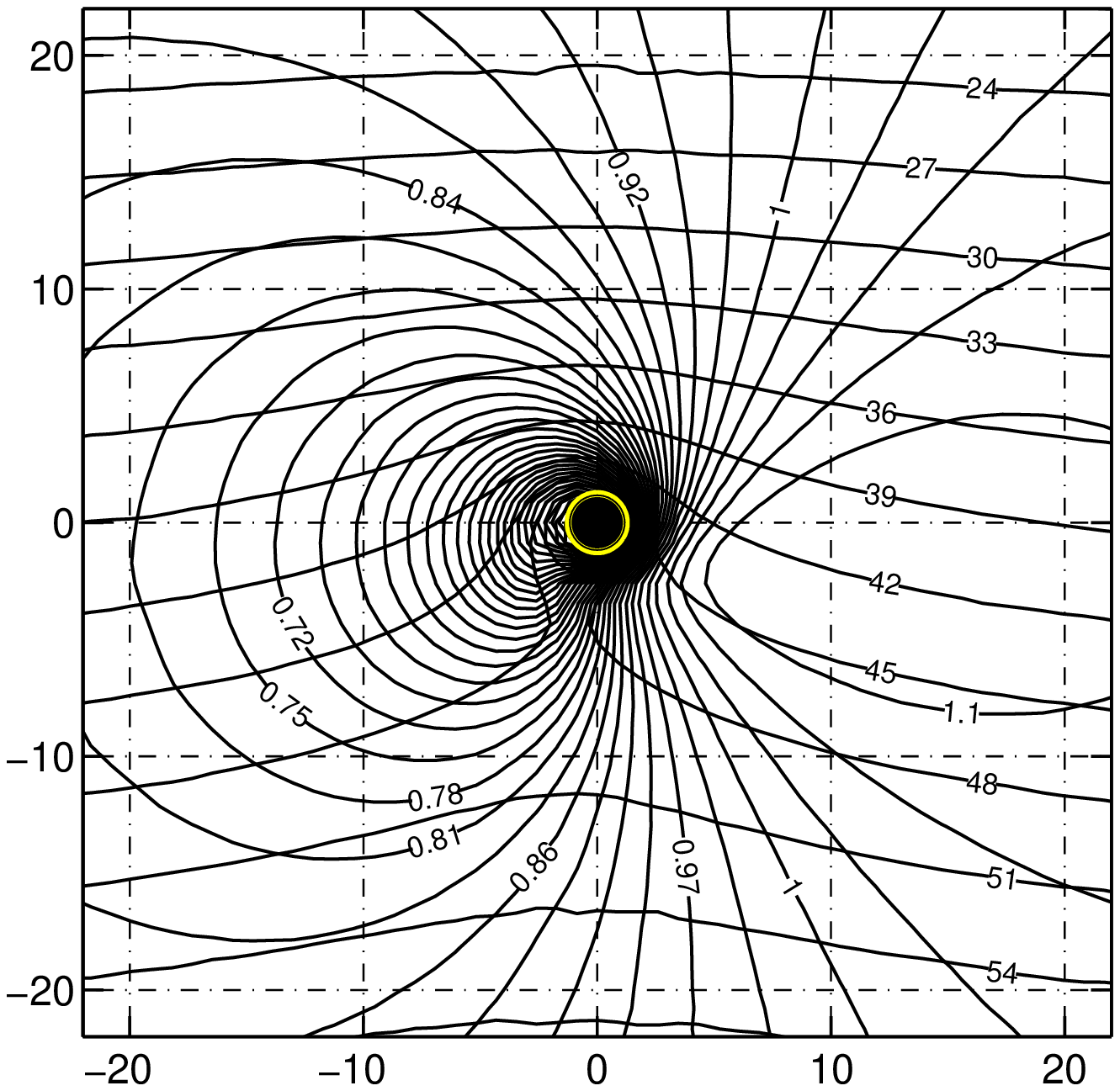}}
\vspace*{2mm}%offset
\epsfxsize=0.37\hsize\makebox[0.42\hsize]{\hspace*{10ex}\epsfbox{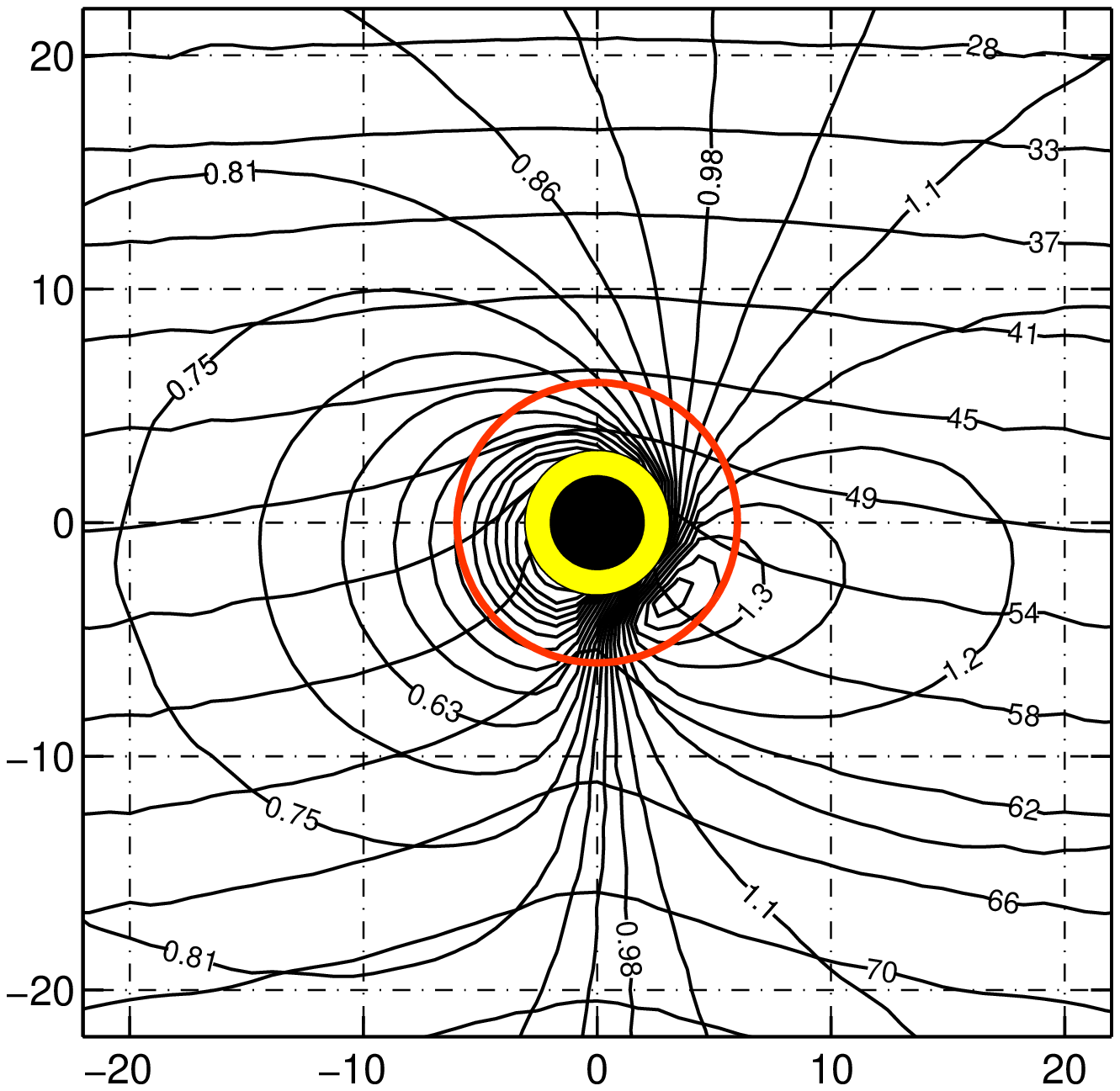}}
\hfill
\epsfxsize=0.37\hsize\makebox[0.42\hsize]{\hspace*{-10ex}\epsfbox{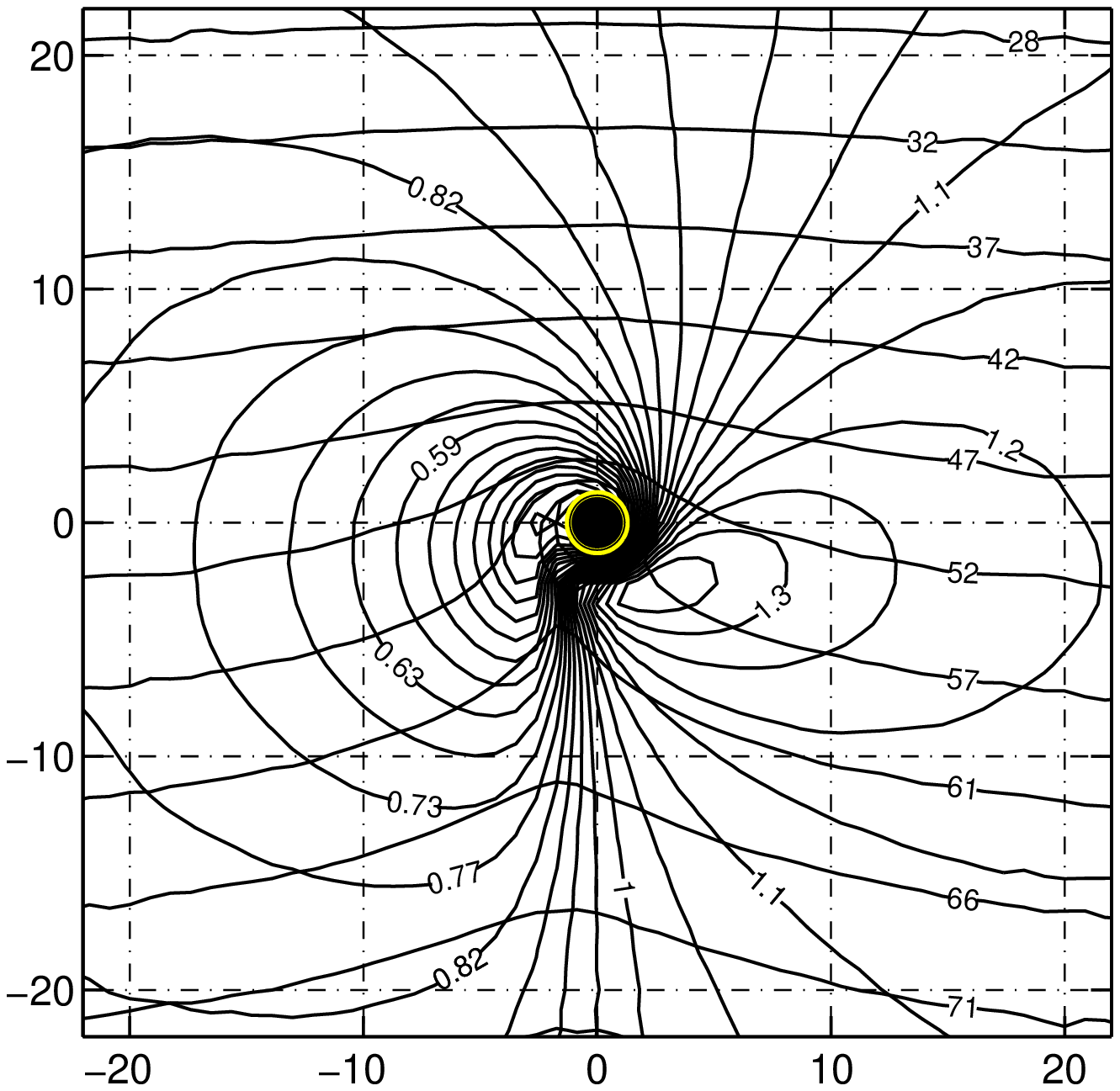}}
\protect\par
\vspace*{-0mm}%offset
\protect\par
{\begin{verse}\footnotesize Fig.~\protect\thefC.\hspace{4pt}~Isocontours
of constant light-travel time $\protect\lag$ and of redshift function
$g$ are shown in equatorial plane of the Kerr black hole. The contours
are shown for a non-rotating black hole ($\protect\astar=0$, left
panels), and for a maximally rotating hole ($\protect\astar=1$, right
panels). Observer inclination is, from top to bottom,
$\protect\thetao=20\protect\dg$, $\protect\thetao=50\protect\dg$, and
$\protect\thetao=80\protect\dg$. For $\astar=0$, three circles are
plotted with radii of the horizon, of the circular photon orbit, and of
the marginally stable orbit. The circles coincide with each other if
$\astar=1$ (corotating orbits are assumed).
\end{verse}}
\end{figure*}%.........................................................

\subsection{Light-Travel Time Near a Black Hole}%---------------------
Now we concentrate our attention on the consequences of
finite light-travel time upon line variations.

One can write an order-of-magnitude (Euclidean) estimate of the maximum
light-travel time delay, $\lagmax$, of rays originating in an
equatorial ring (radius $\rme$ centered on the black hole):
$\lagmax(\rme,\thetao)\approx\pi^{-1}r_{\rm{e}}^{-1/2}\sin\thetao$
($\thetao$ is the observer's inclination). Here, $\lagmax$ characterizes
the mutual delay of the signals emitted at different parts of the ring, and
is defined as the span in arrival times of the rays originating at
different $\varphie$.\footnote{Light rays were approximated as straight
lines for the purpose of estimating $\protect\lagmax$ here. For consistency 
with subsequent paragraphs, the standard notation will be adopted for
Kerr spacetime in Boyer--Lindquist coordinates (Misner et al.\ 1973). 
Geometrized units $c=G=1$ are used and all lengths are made
dimensionless by expressing them in units of the typical mass of the
central black hole, $M\approx M_7\equiv10^7M_\odot$.} Usually, $\rme$ is
supposed to be greater than a marginally stable orbit, $\rms$:
$\rms=3\rg$ for a non-rotating black hole with $\astar=0,$ and
$\rms\rightarrow\rg$ for a maximally rotating black hole,
$\astar\rightarrow1$ ($\rg\equiv1+\sqrt{1-\astar^2}$). Light-travel time
can be ignored provided $\lagmax\ll1$. On the other hand, the lags
become increasingly important when the radiation source extends very
close to the black-hole horizon where the resulting situation is
further complicated by the lensing effect.

The geometrical time lag can be roughly characterized by $\lagmax$, which
indicates whether the Euclidean formula estimates the correct value of the
light-travel time with an acceptable precision, or whether relativistic
effects modify the actual delays significantly. Figure~\thefB\ gives
$\lagmax$ for a source located at $\rme\geq\rms$. The solid curves represent 
the values of the delay computed in Kerr spacetime, while the dotted lines
(slope $-1/2$) show the approximation. Relativistic corrections 
become increasingly important for $\rme/\rg\ltasim3$, in which case 
$\lagmax$ increases sharply. On the
other hand, the difference between the exact value of $\lagmax$ and
its Euclidean approximation is less than 10\,\% for a source location 
$\gtasim5\rg$.

Figure \thefC\ shows the absolute value of light-travel time difference
$\lag$ between a reference ray (arbitrarily chosen but fixed) and a ray
coming from a given point of emission $\{\rme,\varphie\}$ in the
equatorial plane ($\theta=90\dg)$. In this figure, $\lag$ was calculated
in Kerr metric (no approximation was used here). Given $\astar$ and 
$\thetao$, the values of $\lag$ determine where in the disc the corrections 
on the time lag play an important role. Reference values of $\protect\lag$ can 
be transformed to physical units, as measured by a distant observer, by the
relation $\protect\bar{\protect\lag}\,{\rm{}[s]}\approx10^2M_7\,\protect\lag$.
The equatorial plane is portrayed from above (along the common rotation
axis of the hole and of the disc), and a distant observer is located at the
top (Cartesian $y\rightarrow\infty$; coordinates $x$ and $y$ are
introduced in the disc plane by $x^2+y^2=r_{\rm{e}}^2$). Contours of
constant time lag are almost horizontal far from the hole. A distortion of
the contours from straight lines occurs near to the hole due to a lensing
effect and gravitational delay. (Indeed, the contour lines can even become 
split into two disjoint parts.) Next, the redshift function, $g$, is also 
plotted by another set of contour lines in figure~\thefC. Relativistic effects 
on the time lag and redshift are more important for a maximally rotating black 
hole (right panels) because of the smaller $\rms$. 

\begin{Fv}{\thefD}
{0pc \epsfxsize=0.475\hsize\mbox{\epsfbox{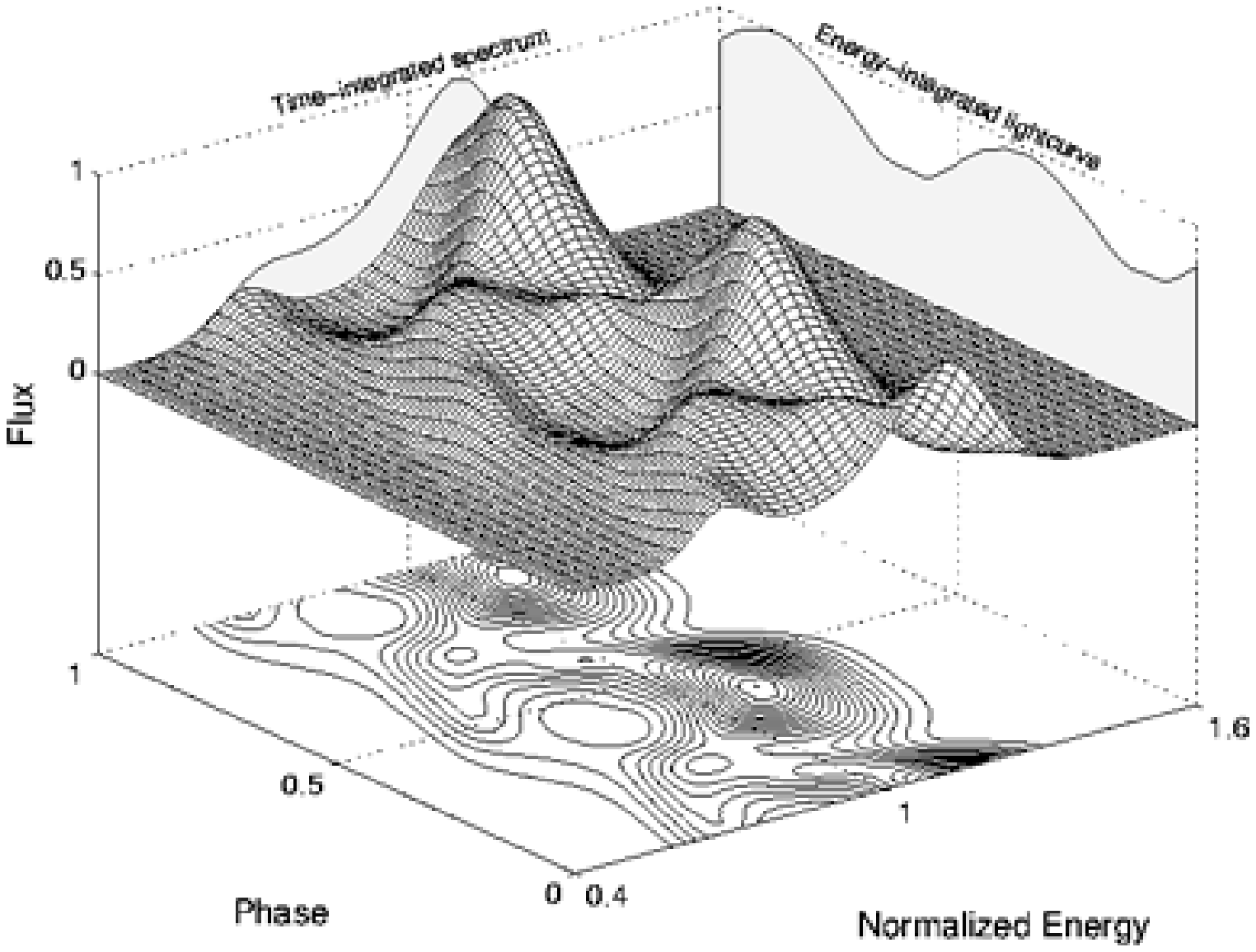}\hfill\
 \epsfxsize=0.475\hsize\epsfbox{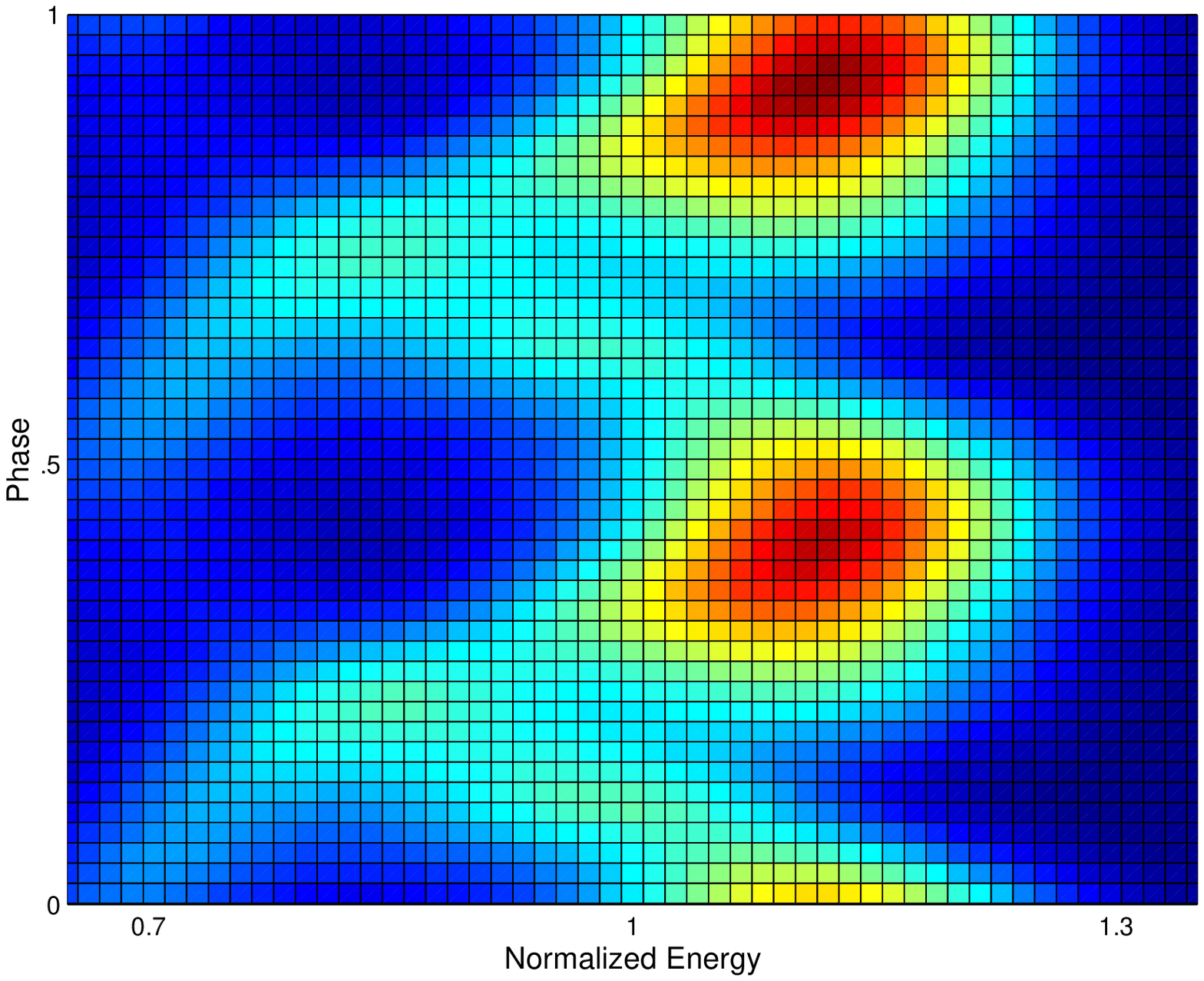}}}{Example of the dynamical
spectrum of a line from a spiral wave near a black hole ($a=0$). Left
panel: The predicted flux is background subtracted and normalized to
maximum. The energy-integrated lightcurve and time-integrated spectrum
are also shown. The parameters for this plot were $\alpha=6$, $\beta=18$,
$\gamma=0$, $r_0=13$, $A_{\rm{}m}=1$. The inclination angle was kept fixed
at $50\dg\!$. Right panel: Projection is shown onto normalized energy
vs.\ phase diagram with flux indicated by different levels of shading.}
\end{Fv}%.................................................................

\subsection{Parameters of Spiral-Wave Line Profiles}%----------------------
Spiral-wave structures produce a rich variety of spectra and
light curves, much broader than we can observe in the case of localized
spots. Apart from the black hole angular momentum, $a$, and the
orientation angle of the source with respect to observer, $\thetao$, one
can identify several parameters related to the geometry and orbital
motion of such spirals which determine the spectral variability:

{\small
\hspace*{\fill}\parbox{0.47\textwidth}{
$\bullet$\hspace*{2ex}The non-axisymmetric intrinsic emissivity of the
source is described by equation~(\protect\ref{j0}) with free parameters
($\alpha$, $\beta$, $\gamma$, $r_0$) determining the azimuthal and
radial dependence of emissivity.}\vspace{1pt}

\hspace*{\fill}\parbox{0.47\textwidth}{
$\bullet$\hspace*{2ex}Doppler shifts between the emitted and 
detected radiation increase the centroid energy and intensity of the
rays originating at the approaching side of the source, and vice versa
for the receding side (factor $g^3$). This effect tends to enlarge the
width of the spectral features and to enhance the observed flux at its
high-energy tail.}\vspace{1pt}

\hspace*{\fill}\parbox{0.47\textwidth}{
$\bullet$\hspace*{2ex}Gravitational redshift affects mainly photons
coming from the inner parts of the source. This effect decreases
the centroid energy of the line.}\vspace{1pt}

\hspace*{\fill}\parbox{0.47\textwidth}{
$\bullet$\hspace*{2ex}Light-travel time determines, along with the shape
of the spiral, which photons are received at the same instant of time.
We remark that contours of constant time delay are deformed near the
hole. Hence, variability amplitude can be much accentuated if the spiral
coincides with one of the contours at certain orbital phase. This effect
influences the flux from some patterns under suitable
orientation.}\vspace{1pt}

\hspace*{\fill}\parbox{0.47\textwidth}{
$\bullet$\hspace*{2ex}Gravitational lensing enhances the flux from the
far side of the source (the part of the source which is at the upper
conjunction with the hole). This tends to increase the observed flux
around the centroid energy.}\vspace{5pt}
}

All parameters of the intrinsic emissivity and of the black hole
spacetime enter in computations of the individual ray trajectories which
form the resulting radiation flux at the detector. We find that the
dependence of observed profiles on $\gamma$ is rather weak for a
typical choice of $0\ltasim\gamma\ltasim2$. One should be aware of other
implicit parameters which were kept fixed in the present model:
the inner radius of the spiral pattern (here $\rms$), the
(small) width of the intrinsic Gaussian local line profile, the number
of spiral patterns (here $n=2$), and duration of the flare
($>t_{\rm{}k}$). In all of these computations, for the sake of simplicity
the radiation was emitted isotropically in the rest frame of the gaseous
medium.

%%%%%%%%%%%%%%%%%%%%%%%%%%%%%%%%%%%%%%%%%%%%%%%%%%%%%%%%%%%%%%%%%%%%%
\section{Results}
A typical line profile of a double-armed spiral is shown in figure~\thefD\ 
where the observed spectral feature is plotted as a function of the observed 
energy and orbital phase. The variation in the spectrum 
is shown over the whole revolution of the pattern by $2\pi$ in azimuth (as
the normalized phase ranges from 0 to 1), while energy axis is
normalized to the local emission energy, $\nue$. Contours of the
dynamical spectrum are shown by projecting the surface plot onto
the plane of energy vs.\ phase. Here, one can clearly identify that the
contribution of the spiral alone is repeated twice in each of the
subplots (because we considered double arms), but the total signal does
not bear this property because of the decaying flare contribution. Next,
figures \thefE--\thefF\ show these contours for different model
parameters. The contours are plotted with uniform spacings between the
maximum flux and zero flux of the signal from which the underlying
continuum is subtracted.

\subsection{The Predicted Form of Observed Profiles}%-------------------
\label{parameters}
The parameter space has been explored systematically in
the following manner.

{\small
\hspace*{\fill}\parbox{0.47\textwidth}{
$\bullet$\hspace*{2ex}Black-hole angular momentum: $\astar=0$ (columns
denoted by letters {\sf{a}}, {\sf{c}}, {\sf{e}}, {\sf{g}}), and
$\astar=1$ ({\sf{b}}, {\sf{d}}, {\sf{f}}, {\sf{h}}). The two cases
correspond to a non-rotating and to maximally rotating hole,
respectively.}\vspace{1pt}

\hspace*{\fill}\parbox{0.47\textwidth}{
$\bullet$\hspace*{2ex}Observer inclination: $\thetao=20\dg$ (the rows
denoted by roman numbers {\sf{i}}--{\sf{iii}}), $\thetao=50\dg$
({\sf{iv}}--{\sf{vi}}), and $\thetao=80\dg$ ({\sf{vii}}--{\sf{ix}}). The
disc plane is $\thetao=90\dg$.}\vspace{1pt}

\hspace*{\fill}\parbox{0.47\textwidth}{
$\bullet$\hspace*{2ex}The outer edge of the spiral pattern: $r_0=7$
({\sf{i}}, {\sf{iv}}, {\sf{vii}}), $r_0=13$ ({\sf{ii}}, {\sf{v}},
{\sf{viii}}), and $r_0=20$ ({\sf{iii}}, {\sf{vi}}, {\sf{ix}}). Units of
$GM/c^2$ are used.}\vspace{1pt}

\hspace*{\fill}\parbox{0.47\textwidth}{
$\bullet$\hspace*{2ex}The form of spiral pattern, as defined by its
pitch angle $\arctan\alpha$ and contrast $\beta$  in eq.~(\ref{j0}):
$\alpha=3$, $\beta=8$ ({\sf{a}}, {\sf{b}}); $\alpha=\sqrt{3}$, $\beta=8$
({\sf{c}}, {\sf{d}}); $\alpha=6$, $\beta=8$ ({\sf{e}}, {\sf{f}});
$\alpha=6$, $\beta=18$ ({\sf{g}}, {\sf{h}}). Large values of $\beta$
correspond to well-defined spirals (with small width and large
contrast), while large values of $\alpha$ define tightly wound spirals
(corresponding to high Mach number in the disc medium).}\vspace{1pt}

\hspace*{\fill}\parbox{0.47\textwidth}{
$\bullet$\hspace*{2ex}A contribution of the flare to the line in terms
of its duration parameter, $q=2$ (units of the orbital period), and the
initial flux, $A_{\rm{}m}$. The average ratio of the flare-to-spiral
intrinsic emissivity in the centre of the line was taken as
$A_{\rm{}m}=1:2^4$ for a weak flare (figure~\thefE) and $A_{\rm{}m}=1:1$
for a strong one (figure~\thefF).}\vspace{5pt}
}

In each of the sub-figures the spectra were computed with sufficient
resolution in energy and phase ($10^2\times10^2$ grid) by the
ray-tracing code (Martocchia 2000).

One can recognize several characteristic features of the dynamical
spectra. The flare is easily distinguished in figure~\thefF\ where it was
assumed to be more intense with respect to the spiral pattern than in
figure~\thefE. However, even in the latter situation the Doppler boosted
radiation of the flare dominates over the extended spiral when the disc
is viewed at large inclination, and especially if the whole source is
near to the horizon ($r_0\approx7$; see the row {\sf{vii}}). Next, one can
recognize the effect of gravitational redshift if the source is seen
almost along the rotation axis ({\sf{i}}--{\sf{iii}}), whereas this effect
turns out to be less visible in very broad spectral features
corresponding to large inclinations ({\sf{vii}}--{\sf{ix}}).

%\begin{Fv}{\thefE}{0pc \epsfxsize=\hsize\epsfbox{prof10.eps}}
\begin{figure*}
\vspace*{-1mm}
\hfill
\epsfxsize=0.90\hsize\epsfbox{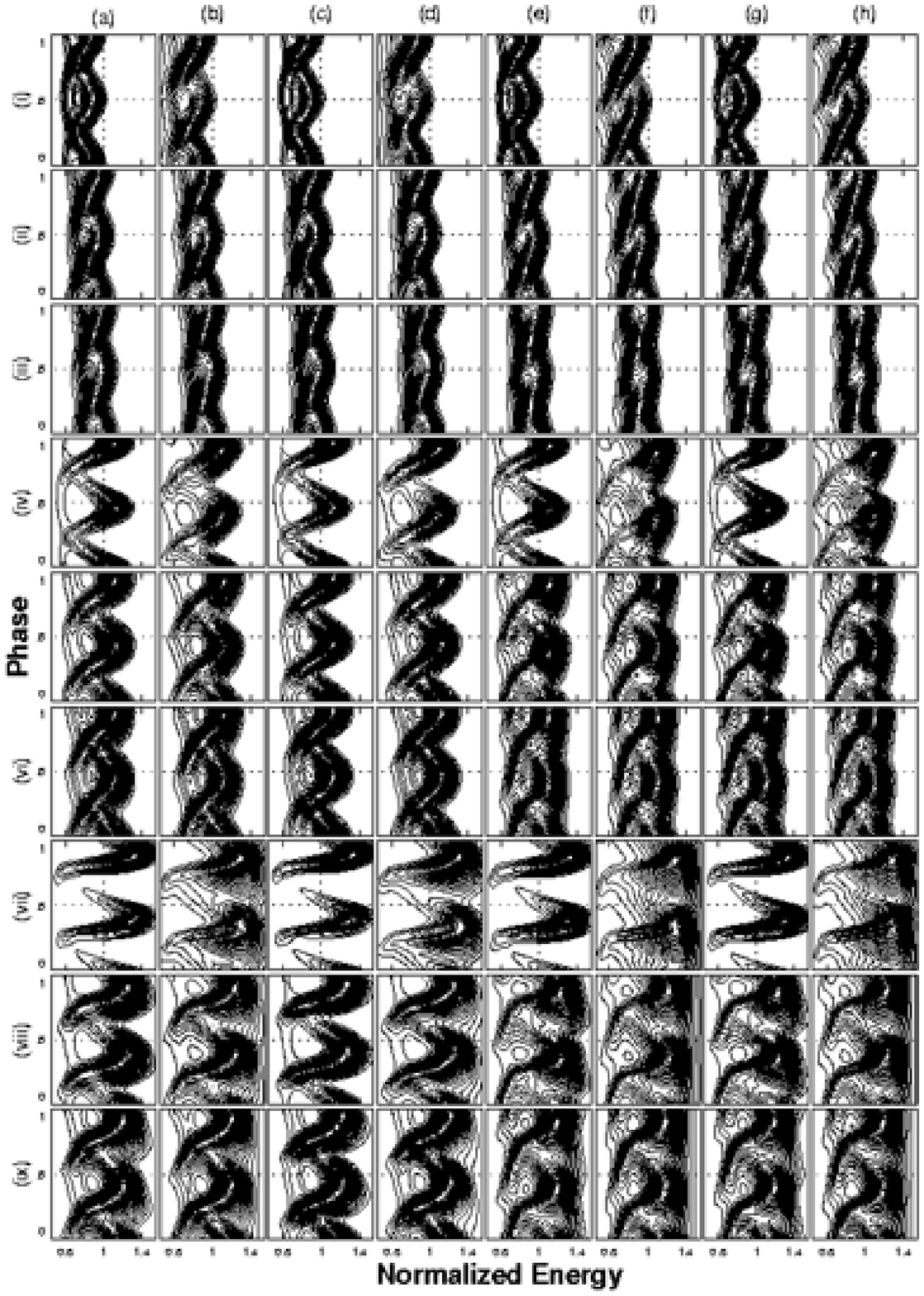}
\hfill
\vspace*{-1mm}%offset
{\begin{verse}\footnotesize Fig.~\protect\thefE.\hspace{4pt}~{Dynamical
spectra of the two-arm pattern with a slowly decaying flare. Different
cases correspond to the form of the spiral, rotation parameter of the
black hole, and observer's view angle (see the text for details). Here,
the spiral dominates over the flare.}
\end{verse}}
\end{figure*}%...........................................................
%\end{Fv}

%\begin{Fv}{\thefF}{0pc \epsfxsize=\hsize\epsfbox{prof05.eps}}
\begin{figure*}
\vspace*{-1mm}
\hfill
\epsfxsize=0.90\hsize\epsfbox{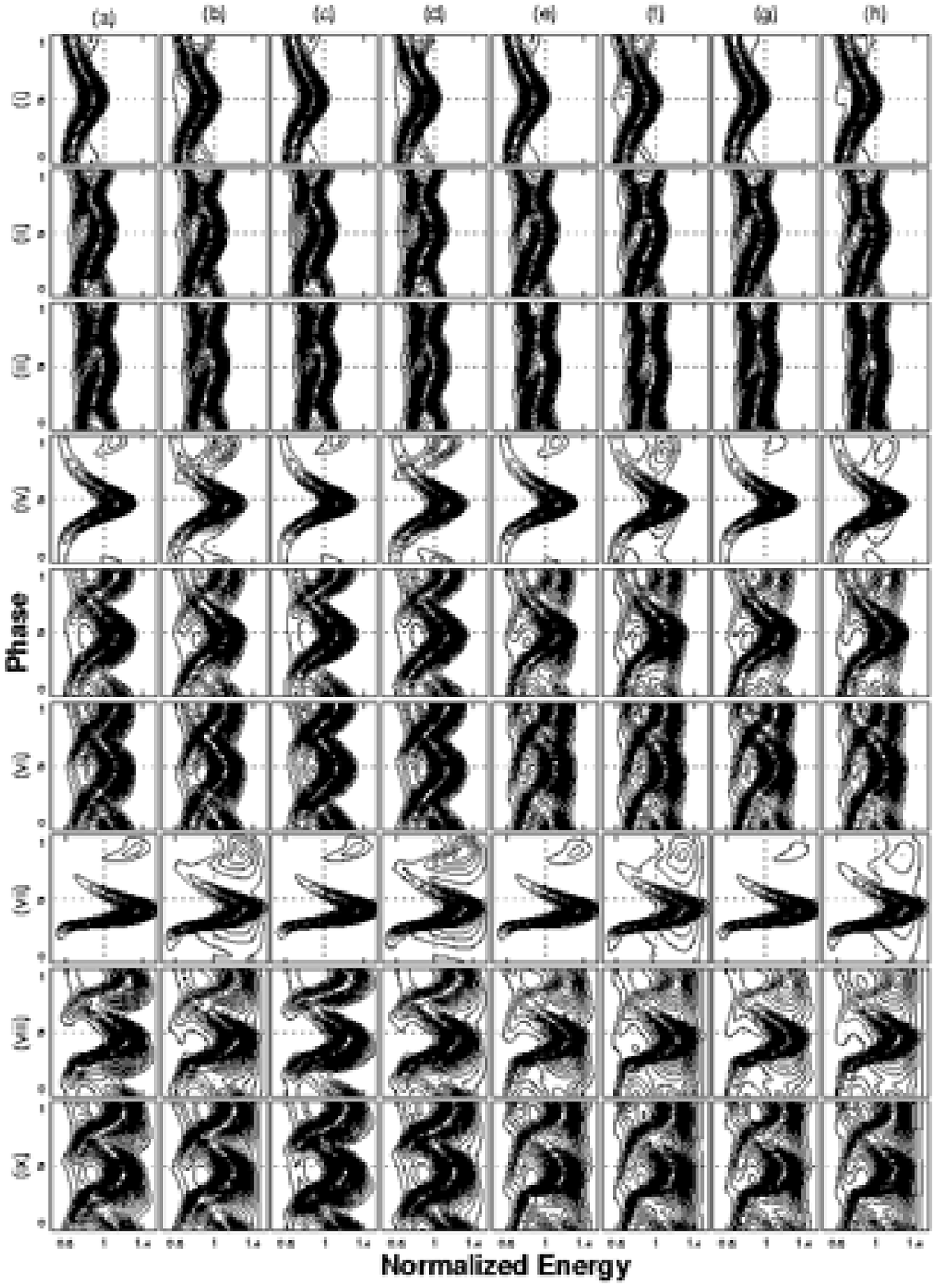}
\hfill
\vspace*{-1mm}%offset
{\begin{verse}\footnotesize Fig.~\protect\thefF.\hspace{4pt}~{Same 
as in figure~\protect\thefE, but now the flare is assumed to be
more pronounced. This represents an intermediate situation between a
pure spiral and the orbiting spot model.}
\end{verse}}
\end{figure*}%............................................................
%\end{Fv}

\subsection{The Case of MCG$-$6--30--15}%.................................
While the two plates in figures \thefE--\thefF\ has given us a
general feeling for the influence of different parameters, a suitable
($\chi^2$ minimizing) routine is used in practice for fitting the
computed profiles to data.

Figure \thefG\ shows the best fits to three iron-line profiles in
MCG$-$6--30--15 (we use exactly the same data as defined by Iwasawa et al.\
1996). These correspond to the spectral states of the source during its 1994
observation with the ASCA satellite. Here, we do not intend to exactly 
mimic what is going on in the source, because this appears to be premature
with the available data. It is, however, possible to constrain the distance
from the centre at which the line could originate, depending on the 
angular momentum of the hole and inclination of the observer.

The curves shown in the right panels are those to which the $\chi^2$ 
minimizing routine converges in the parameter space (as described in 
subsection \ref{parameters}). The reduced $\chi^2$ is (from top to bottom):
$\chi^2/{\sf{d.o.f.}}=72/53$, $40/55$, $60/40$ for the three phases,
respectively. Here, $\astar=0$, $\thetao=20\dg$, $r_0=13$, $\alpha=6$,
$\beta=18$, $A_{\rm{}m}=1/16$. We thus find that the data are well
fitted by the double arm at moderate inclination angles. In such a case,
the dependence of the observed profile on $\astar$ is negligible because
a substantial part of the line flux originates sufficiently above $\rms$.

\begin{fv}{\thefG}{0pc \epsfxsize=\hsize\epsfbox{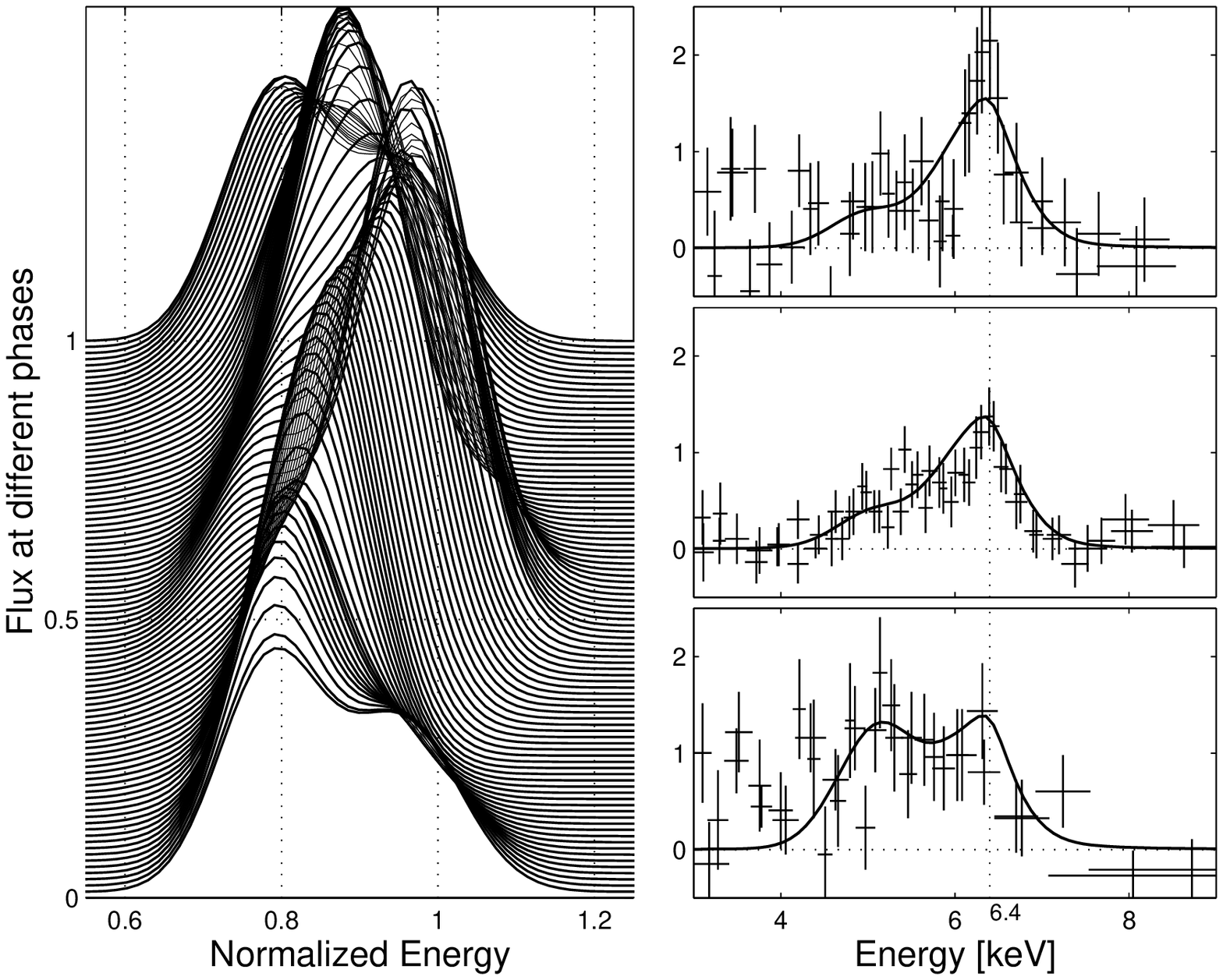}}
{Left: Time variation of the best-fit profile during the entire
revolution of the spiral. Right: The computed profile is plotted after
averaging to three different phases. The feature observed in
MCG$-$6--30--15 (Iwasawa et al.\ 1996) is overlayed with an appropriate 
phase shift.}
\end{fv}%................................................................

%%%%%%%%%%%%%%%%%%%%%%%%%%%%%%%%%%%%%%%%%%%%%%%%%%%%%%%%%%%%%%%%%%%%%%
\section{Discussion}
In the present work we did not intend to ameliorate the fits of 
MCG$-$6--30--15 data further. It is, however, worth noticing that even the
adopted, rather simple scheme produces acceptable fits and the required,
reddened line centroid also with small values of $a$.
One can thus address the question if those redshifted and variable
profiles of the iron line could be achieved without the necessity of
invoking a large mass ($>10^7M_\odot$) of the central object and/or
its fast rotation, which have recently been put in certain doubt by several
people. It is thus legitimate to consider all different viable options
because various effects contribute to the line formation, and the
interpretation of observed profiles is not straightforward.\footnote{Weaver 
and Yaqoob (1998) and Reynolds and Begelman (1997)
proposed alternative schemes which do not involve fast rotation of the
black hole. Nowak and Chiang (2000) discussed the possible caveats in
interpretations of the present time-resolved spectra, and they concluded
that the central mass is about $10^6M_\odot$, much less than previously
considered. A rather small mass, however, requires efficient emission,
which also seems to pose a problem, so the issue has not been settled
yet.} A self-consistent interpretation of the profiles cannot be achieved
without regard to the underlying continuum component where it is crucial
to distinguish between line and continuum photons.

Our discussion is relevant for X-ray variability of some spectral
states observed in AGN and GBHCs; it is quite obvious, however, that
only a part of the complex variability behaviour could be attributed to
motion of large scale structures, such as spiral waves. The model is
complementary to the discussion of instant off-axis flares inducing the
observed line variations, which enable one to restrict the geometry of
the source by means of the reverberation technique. Proper evaluation of
the light-travel time from different parts of the source is important in
both schemes.
         
%%%%%%%%%%%%%%%%%%%%%%%%%%%%%%%%%%%%%%%%%%%%%%%%%%%%%%%%%%%%%%%%%%%%%
\section{Conclusions}
We computed variable line profiles due to non-axisymmetric
spiral-type patterns near a Kerr black hole. We adopted several
simplifying assumptions and constrained some of the parameters to fixed
values. Yet, we could show that a big variety of line profiles are
reproduced and the data of MCG$-$6--30--15 which we used for illustration
purposes can be modeled without the necessity of $\astar\rightarrow1$,
although such a possibility is not rejected. Because the existence of spiral
features is expected on the basis of theoretical models, our
discussion seems to be better substantiated than data fitting by ad hoc
functions. Also, the adopted approach uses computed profiles, rather
than being restricted to integral quantities (centroid energy, line
width, etc.); it can thus provide robust results once better-resolved
data become available.

Our present scheme can be considered as complementary to the picture of
Reynolds et al.\ (1999) and Ruszkowski
(2000), who examined off-axis flares above the disc surface as the origin
of line variations, and to Vaughan and Edelson (2001), who addressed the
problem of the link between line and continuum variations. Naturally, 
the phenomenological description of the source will have to be supplemented 
by a time-dependent physical model of X-ray reprocessing.

Although we applied our computations only to the case of MCG$-$6--30--15,
there is a more general interest in the problem of spectral features
from inner parts of gaseous discs endowed with spirals. Compact binaries
with matter overflow represent another class of objects where our
computations could be useful. Patterns like those discussed in this
paper can be introduced in order to explain asymmetric lines and
variability observed in UV/optical spectral bands from the broad-line
region (Dumont, Collin-Soufrin 1990; Corbin 1997), and periodicities
reported in the optical light curve of the blazar OJ\,287 (Takalo 1994;
Villata et al.\ 1998). In these cases, however, the radiation is emitted
at distances exceeding $10^3$ gravitational radii, so that the general
relativity calculations are unnecessary and the computations can be
performed in a much faster way.

Even though the results cannot be conclusive with the present-day
resolution, there is evident potentiality of this approach with planned
missions, like Constellation-X and XEUS. The model of spirals illustrates
the opportunity of tracing the surface structures on accretion discs.

We acknowledge support from the grants GACR 205/00/1685 and 202/99/0261.
V\,K thanks for hospitality of the International School for Advanced
Studies (Trieste) and Observatoire de Paris-Meudon.

\normalsize
  
%\begin{thebibliography}{}
\section*{References}%%%%%%%%%%%%%%%%%%%%%%%%%%%%%%%%%%%%%%%%%%%%%%%%%
\small
\re%\bibitem{ABLZ91}
 Abramowicz, M.\,A., Bao,~G., Lanza,~A., \& Zhang, X.-H. 1991, A\&A, 245, 454
\re%\bibitem{ALSS92}
 Abramowicz, M.\,A., Lanza,~A., Spiegel, E.\,A., \& Szus\-z\-kie\-wicz,~E.
 1992, Nature, 356, 41
\re%\bibitem{AC00}
 Abrassart,~A., \& Czerny,~B. 2000, A\&A, 356, 475%(astro-ph/0001515)
\re%\bibitem{AW95}
 Adams, F.\,C., \& Watkins,~R. 1995, ApJ, 451, 314
\re%\bibitem{BW99}
 Bao,~G., \& Wiita,~P. 1999, ApJ, 519, 80
\re%\bibitem{BK82}
 Blandford, F.\,D., \& McKee, C.\,F. 1982, ApJ, 255, 419
\re%\bibitem{CW93}
 Chakrabarti, S.\,K., \& Wiita, P.\,J. 1993, A\&A, 271, 216
\re%\bibitem{CW94}
 Chakrabarti, S.\,K., \& Wiita, P.\,J. 1994, ApJ, 434, 518
\re%\bibitem{CH90}
 Chen, K., \& Halpern, J.\,P. 1990, ApJ, 354, L1
\re%\bibitem{C92}
 Coppi, P.\,S. 1992, MNRAS, 258, 657
\re%\bibitem{C97}
 Corbin, M.\,R. 1997, ApJ, 485, 517
%\re%\bibitem{DL00}
%Dabrowski,~Y., \& Lasenby, A.\,N. 2001, MNRAS, submitted (astro-ph/0005020)
\re%\bibitem{DN01}
 Done, C., \& Nayakshin, S. 2001, ApJ, 546, 419%(astro-ph/0008292)
\re%\bibitem{DC90}
 Dumont, A.-M., \& Collin-Soufrin,~S. 1990, A\&A Suppl, 229, 313
\re%\bibitem{DAC00}
 Dumont, A.-M., Abrassart,~A., \& Collin,~S. 2000, A\&A, 357,
 823%(astro-ph/0003220)
\re%\bibitem{E98}
 Eracleous,~M. 1998, Adv.\ Space Res., 21, 33
\re%\bibitem{FIRY00}
 Fabian, A.\,C., Iwasawa,~K., Reynolds, C.\,S., \& Young, A.\,J. 2000,
 PASP, 112, 1145%(astro-ph/0004366)
\re%\bibitem{FCFC97}
 Fanton,~C., Calvani,~M., de~Felice,~F., \& \v{C}ade\v{z},~A.
 1997, PASJ, 49, 159
\re%\bibitem{GRV79}
 Galeev, A.\,A., Rosner,~R., \& Vaiana, G.\,S. 1979, ApJ, 229, 318
\re%\bibitem{GT79}
 Goldreich, P., \& Tremaine,~S. 1979,  ApJ, 233, 857
\re%\bibitem{HMG94}
 Haardt,~F., Maraschi,~L., \& Ghisellini,~G. 1994, ApJ, 432, L95
\re%\bibitem{IFYIM99}
 Iwasawa,~K., Fabian, A.\,C., Young, A.\,J., Inoue,~H., \& Matsumoto,~C.
 1999, MNRAS, 306, L19
\re%\bibitem{I96}
 Iwasawa,~K., Fabian, A.\,C., Reynolds, C.\,S., Nandra,~K., Otani, C., 
 Inoue, H., Hayashida, K., Brandt, W.\,N.\ et al. 1996, MNRAS, 282, 1038
\re%\bibitem{K87}
 Karas, V. 1997, MNRAS, 288, 12
\re
 Karas, V., Czerny B., Abrassart A., \& Abramowicz M.\,A. 2000, MNRAS,
 318, 547
\re%\bibitem{KFM98}
 Kato,~S., Fukue,~J., \& Mineshige,~S. 1998, Black-Hole Accretion Disks
 (Kyoto: Kyoto University Press)
\re%\bibitem{KMMMS00}
 Kawaguchi,~T., Mineshige,~S., Machida,~M., Matsumoto,~R., \& Shibata,~K.
 2000, PASJ, 52, L1
\re%\bibitem{K99}
 Krolik, J.\,H. 1999, Active Galactic Nuclei: From the Central Black Hole
 to the Galactic Environment (Princeton: Princeton University Press)
\re%\bibitem{KY99}
 Kuo,~C.-L., \& Yuan,~C. 1999, ApJ, 512, 79
\re%\bibitem{LMB}
 Lanzafame, G., Maravigna, F., \& Belvedere, G. 2000, PASJ, 52, 515
\re%\bibitem{LP93}
 Lawrence,~A., \& Papadakis,~I. 1993, ApJ, 414, L85
\re
 Lee,~E., \& Goodman,~J. 2000, MNRAS, 308, 984
\re%\bibitem{MJ00}
 Malzac,~J., \& Jourdain,~E. 2000, A\&A, 359, 843%(astro-ph/0005523)
\re%\bibitem{MW93}
 Mangalam,~A.\,V., \& Wiita, P.\,J. 1993, ApJ, 406, 420
\re%\bibitem{MM98}
 Maoz,~E., \& McKee, C.\,F. 1998, ApJ, 494, 218
\re%\bibitem{MH88}
 Marsh,~T.\,R., \& Horne,~K. 1988, MNRAS, 235, 269
\re%\bibitem{M00}
 Martocchia,~A. 2000, X-ray Spectral Signatures of Accreting Black Holes,
 Ph.D.\ Thesis, SISSA Trieste
\re%\bibitem{MKM00}
 Martocchia,~A., Karas,~V., \& Matt,~G. 2000, MNRAS, 312, 817
\re%\bibitem{MM96}
 Martocchia,~A., \& Matt,~G. 1996, MNRAS, 282, L53
\re
 Misner, C.\,W., Thorne, K.\,S., \& Wheeler, J.\,A. 1973, Gravitation
 (New York: W.\,H.\ Freeman and Co.)
\re%\bibitem{MON94}
 Mineshige,~S., Ouchi, N.\,B., \& Nishimori,~H. 1994, PASJ, 46, 97
\re%\bibitem{MAFW99}
 Murray, J.\,R., Armitage, P.\,J., Ferrario,~L., \& Wickramasinghe, D.\,T.
 1999, MNRAS, 302, 189
\re%\bibitem{N98}
 Nayakshin, S.\,V. 1998, Physics of Accretion Disks with Magnetic Flares,
 Ph.D.\ Thesis, University of Arizona%(astro-ph/9811061)
\re%\bibitem{NKK99}
 Nayakshin,~S., Kazanas,~D., \& Kallman, T.\,R. 2000, ApJ,
 537, 798%(astro-ph/9909359)
\re%\bibitem{NC00}
 Nowak, M.\,A., \& Chiang,~J. 2000, ApJ, 531, L13
\re%\bibitem{RB94}
 Rauch,~K., \& Blandford, R.\,D. 1994, ApJ, 421, 46
\re%\bibitem{RB97}
 Reynolds, C.\,S., \& Begelman, M.\,C. 1997, ApJ, 488, 109
\re%\bibitem{RYBF99}
 Reynolds, C.\,S., Young, A.\,J., Begelman, M.\,C., \& Fabian, A.\,C. 1999,
 ApJ, 514, 164
\re%\bibitem{RCB84}
 Rickett, B.\,J., Coles, W.\,A., \& Bourgois,~G. 1984, A\&A, 134, 390
\re%\bibitem{RS93}
 R\'o\.zyczka,~M., \& Spruit, H.\,C. 1993, ApJ, 417, 677
\re%\bibitem{R00}
 Ruszkowski,~M. 2000, MNRAS, 315,~1
\re%\bibitem{SFK94}
 Sanbuichi,~K., Fukue,~J., \& Kojima,~Y. 1994, PASJ, 46, 605
\re%\bibitem{SH76}
 Sanders, R.\,H., \& Huntley, J.\,M. 1976, ApJ 209, 53
\re%\bibitem{SMH86}
 Sawada,~K., Matsuda,~T., \& Hachisu,~I. 1986, MNRAS, 219, 75
\re%\bibitem{SS99}
 Steeghs,~D., \& Stehle,~R. 1999, MNRAS, 307, 99
\re%\bibitem{THSP90}
 Tagger,~M., Henriksen, R.\,N., Sygnet, J.\,F., \& Pellat,~R. 1990,
 ApJ, 353, 654
\re%\bibitem{TP99}
 Tagger,~M., \& Pellat,~R. 1999, A\&A, 349, 1003
\re%\bibitem {T94}
 Takalo,~L. 1994, Vistas Astron., 38, 77
\re%\bibitem{T96}
 Taylor, J.\,A. 1996, ApJ, 470, 269
\re%\bibitem{UMU97}
 Ulrich, M.-H., Maraschi,~L., \& Urry, C.\,M. 1997, ARA\&A, 35, 445
\re
 Vaughan, S., \& Edelson, R. 2001, ApJ, preprint%(astro-ph/0010274)
\re%\bibitem{VRST98}
 Villata,~M., Raiteri, C.\,M., Sillanp\"a\"a,~A., \& Takalo, L.\,O. 1998,
 MNRAS, 293, L13
\re%\bibitem{W94}
 Wada,~K. 1994, PASJ, 46, 165
\re%\bibitem{WY98}
 Weaver, K.\,A., \& Yaqoob,~T. 1998, ApJ, 502, L139
\re%\bibitem{WBBMS98}
 Wolf,~S., Barwig,~H., Bobinger,~A., Mantel, K.-H., \& \v{S}imi\'c,~D. 1998,
 A\&A, 332, 984
\re%\bibitem{YMW97}
 Yonehara, A., Mineshige, S., \& Welsh, W.\,F. 1997, ApJ, 486, 388
%\end{thebibliography}

\label{last}
\end{document}